\documentclass[12pt]{article}
\usepackage{times}
\usepackage{geometry}
\usepackage{subfigure}
\usepackage[utf8]{inputenc}
\usepackage{enumitem,amssymb}
\usepackage{natbib}
\usepackage{ragged2e}
\usepackage{color}
\newlist{thematic}{itemize}{8}
\setlist[thematic]{label=$\square$}
\usepackage{pifont}
\usepackage{graphics}
\usepackage{rotating}
\usepackage{wrapfig}
\newcommand{\cmark}{\ding{51}}%
\newcommand{\done}{\rlap{$\square$}{\raisebox{2pt}{\large\hspace{1pt}\cmark}}%
\hspace{-2.5pt}}

\begin{document}
\raggedright
\huge
Astro2020 Science White Paper \linebreak

EUV observations of cool dwarf stars \linebreak
\normalsize

\noindent \textbf{Thematic Areas:} \hspace*{60pt} $\square$ Planetary Systems \hspace*{10pt} $\square$ Star and Planet Formation \hspace*{20pt}\linebreak
$\square$ Formation and Evolution of Compact Objects \hspace*{31pt} $\square$ Cosmology and Fundamental Physics \linebreak
  \done~Stars and Stellar Evolution \hspace*{1pt} $\square$ Resolved Stellar Populations and their Environments \hspace*{40pt} \linebreak
  $\square$    Galaxy Evolution   \hspace*{45pt} $\square$             Multi-Messenger Astronomy and Astrophysics \hspace*{65pt} \linebreak
  
\textbf{Principal Author:}

Name: Allison Youngblood
 \linebreak						
Institution:  NASA Goddard Space Flight Center
 \linebreak
Email: allison.a.youngblood\@nasa.gov
 \linebreak
Phone:  (301) 614-5729
 \linebreak
 
\textbf{Co-authors:}

Jeremy Drake (CfA), James Mason (NASA GSFC), Rachel Osten (STScI, JHU), Meng Jin (Lockheed Martin), Adam Kowalski (University of Colorado), Kevin France (University of Colorado), Brian Fleming (University of Colorado), Joel Allred (NASA GSFC), Ute Amerstorfer (IWF, Graz), Zachory Berta-Thompson (University of Colorado), Vincent Bourrier (University of Geneva), Luca Fossati (IWF, Graz), Cynthia Froning (University of Texas), Cecilia Garraffo (Harvard CfA), Guillaume Gronoff (NASA LRC), Tommi Koskinen (University of Arizona), Herbert Lichtenegger (IWF, Graz)
\medskip

\textbf{Abstract:}
The EUV (100-912 \AA) is a spectral region notoriously difficult to observe due to attenuation by neutral hydrogen gas in the interstellar medium. Despite this, hundreds to thousands of nearby stars of different spectral types and magnetic activity levels are accessible in the EUV range. The EUV probes interesting and complicated regions in the stellar atmosphere like the lower corona and transition region that are inaccessible from other spectral regions. In this white paper we describe how direct EUV observations, which require a dedicated grazing-incidence observatory, cannot yet be accurately substituted with models and theory. Exploring EUV emission from cool dwarf stars in the time domain can make a major contribution to understanding stellar outer atmospheres and magnetism, and offers the clearest path toward detecting coronal mass ejections on stars other than the Sun.


\pagebreak

\section*{The Nature of Stellar EUV Emission}
Dynamo action in cool stars (T$_{eff}$ $\lesssim$ 7500 K) induced by rotation and convection generates hot plasma (10$^4$-10$^7$ K) that becomes trapped within closed magnetic structures, comprising the chromosphere, transition region, and coronal layers of a star's atmosphere (e.g. Vaiana \& Rosner, 1978). X-ray emission from cool stars arises in the corona, which is comprised of optically-thin plasma (10$^6$-10$^7$ K) that is approximately in thermal equilibrium and relatively easy to observe and model. The EUV spectral range exhibits some of this hot optically-thin emission, but a significant fraction originates from the transition region ($\sim$10$^5$ K) and cooler plasma of the corona (10$^{6}$ K), and is more complicated to observe and model.
\medskip

\begin{figure}
    \centering
    \subfigure{
    \includegraphics[width=0.85\textwidth]{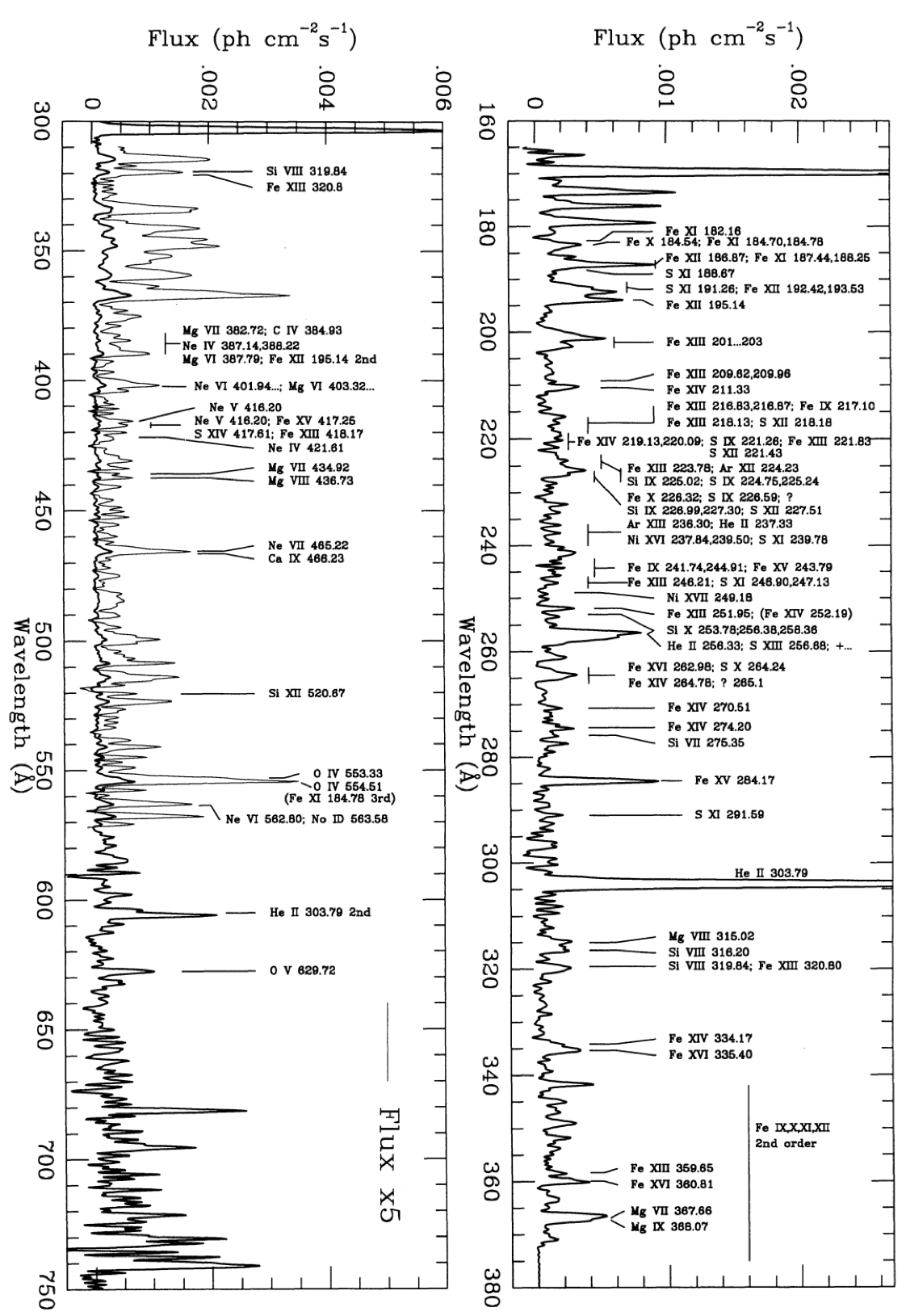}
    }
    \caption{The \textit{EUVE} Medium Wavelength and Long Wavelength spectrometer observations of the nearby F4 subgiant Procyon from a 100ks exposure spanning more than 3 days. Prominent spectral lines are identified. Attenuation from neutral hydrogen gas in the local interstellar medium prohibit access to photons from $\gtrsim$ 650 \AA, even for bright nearby stars. From \cite{Drake1995}. }
    \label{fig:EUVE_Procyon}
\end{figure}

Solar and stellar EUV spectra are characterized by emission lines from abundant ionized chemical elements superimposed on a continuum (Figure~\ref{fig:EUVE_Procyon}; e.g. \citealt{Doschek1991,Feldman1992,Drake1999}). Emission lines are formed by collisional excitation and subsequent decay; the continua are produced in recombination free-bound transitions.
\medskip 

The {\it Extreme Ultraviolet Exporer (EUVE)} remains the only observatory to have made extensive spectroscopic observations of cosmic sources in the EUV range. The {\it EUVE} spectrometers had an effective area of $\sim$1~cm$^2$, necessitating days-long exposure times generally resulting in low signal-to-noise spectra and limited time resolution. Despite this, {\it EUVE} made remarkable breakthroughs, including opening up to stellar physics the wealth of coronal plasma diagnostics enjoyed by solar physicists for decades. A new large-area EUV spectroscopy mission could make major breakthroughs in the study of stellar outer atmospheres; in turn, these observations will be crucial for understanding the effects of stellar energetic emission on exoplanets (see white paper titled ``EUV influences on exoplanet atmospheric stability and evolution'' led by A. Youngblood).

\section*{Stellar EUV Emission Cannot be Inferred from Observations at Other Wavelengths}

Stellar models that self-consistently treat the chromosphere, transition region, and corona are in their infancy \citep{Fontenla2016,Peacock2019}, while the fundamental physics of coronal heating remains poorly understood. Differential emission measure (DEM) models can generate EUV spectra under the assumption of thermal equilibrium and optically thin conditions with three parameters: temperature, density, and chemical composition. Constraining these three parameters requires a substantial amount of data describing the collisional excitation and ionization processes involved. Complete DEMs require observations sampling the emission measure of the  gas across a very wide range of temperatures, 10$^4$–10$^8$ K \citep{delZanna2002,Osten2006}. However, the great majority of spectral lines formed in the 10$^5$–10$^6$ K temperature range (with the notable exception of the O VI doublet at FUV wavelengths) can only be observed in the EUV. 
\medskip

Traditional DEM analyses also suffer from other problems:
\begin{itemize}
\item Coronal and transition region emission are strongly variable on a variety of timescales, ranging from seconds (in stellar flares) to years (magnetic cycles) to eons (stellar spindown and decline in magnetic activity). DEMs reconstructed from non-simultaneous observations in different wavelengths---e.g., X-ray and UV---will be subject to large uncertainties due to variability. It is not yet known how the full DEMs of stellar coronae vary over time or during flares.

\item The derivation of the DEM based on observations of spectral lines---an integral inversion---is also a mathematically ill-conditioned problem (e.g. \citealt{Craig1976}). Further constraints are generally required, such as smoothing criteria (see, e.g., the discussion in \citealt{Kashyap1998}), and spectral lines finely sampling the full range of coronal temperatures are needed.
Ill-constrained DEMs can lead to large uncertainties in modeled spectral output.

\item Element abundances in solar and stellar coronae are subject to strong fractionation by mechanisms that are not yet understood, but appear to depend on plasma temperature and spectral type (e.g., the first ionization potential (FIP) effect; \citealt{Drake1997,Laming2015,Wood2018}). Emergent flux scales linearly with chemical abundance, so typical abundance uncertainties of factors of 4 or more translate to uncertainties in line fluxes by a similar amount. 
\end{itemize}

\section*{Flares in the EUV}
Existing stellar coronal flare analyses have been limited to very large flares on active stars, while favoring the hotter temperature diagnostics in the X-ray region. The bias in temperatures probed by \textit{EUVE} -- mainly hot lines that sample temperatures similar to X-ray lines -- resulted from lack of sensitivity to lower temperature transitions, as well as lack of sensitivity at longer wavelengths. {\it EUVE} flare analyses are limited to a handful of low S/N observations. Filling in the lower temperatures, as well as a range of flare sizes and on a range of different types of stars, is a relatively unexplored parameter regime for stellar flares---an ample discovery space that is also ripe for comparison with the well-sampled solar measurements from \textit{SDO}/EVE, \textit{SDO}/AIA, \textit{SOHO}/EIT, \textit{IRIS}, \textit{GOES}/SUVI, \textit{GOES}/EXIS, and \textit{PROBA-2}/SWAP.
\medskip

There are many open questions regarding the structure of flaring regions on other stars. In particular, the EUV late phase of flares was discovered in Sun-as-a-star measurements made by \textit{SDO}/EVE; in some cases these later phases contain more energy than the main peak of the flare \citep{Woods2014}, which has consequences for the impact on planetary atmospheres. Hours after the main hot flare phase ($\sim$10$^7$ K), a second much longer emission enhancement can occur at cooler coronal temperatures (10$^{6.4}$-10$^{6.8}$ K), which has only been in the EUV for the Sun \citep{Woods2014}.

While there appears to be general agreement about the behavior of the radiative component of stellar flares and solar flares, the role of electron beams and accelerated particles indicates disagreement. The work of \cite{Allred2006} and later \cite{Kowalski2015,Kowalski2017} demonstrates that a solar-like prescription of an electron beam in a realistic M dwarf stellar atmosphere does not reproduce the observed stellar flare characteristics in the blue-optical wavelength ranges. Therefore, while the plasma heating characteristics of small to moderate coronal stellar flares seems to be consistent with solar flares, M dwarf observations are revealing a growing disconnect in the particle acceleration properties. For example, \cite{Froning2019} found that the hot FUV flare continuum from a large M dwarf flare could be reproduced by radiative hydrodynamic models only with the \textit{ad hoc} addition of a hot, dense emitting component not explained by typical electron beam energy fluxes \citep{Kowalski2015}. The EUV spectral region probes the largest swathe of temperature space compared to the X-ray and the FUV, and time-resolved spectroscopy of stellar flares could resolve some of these outstanding issues.

\section*{How can we characterize coronal mass ejections from other stars? A path forward in the EUV}

\begin{figure}
   \begin{center}
   \subfigure{
    \includegraphics[width=0.5\textwidth]{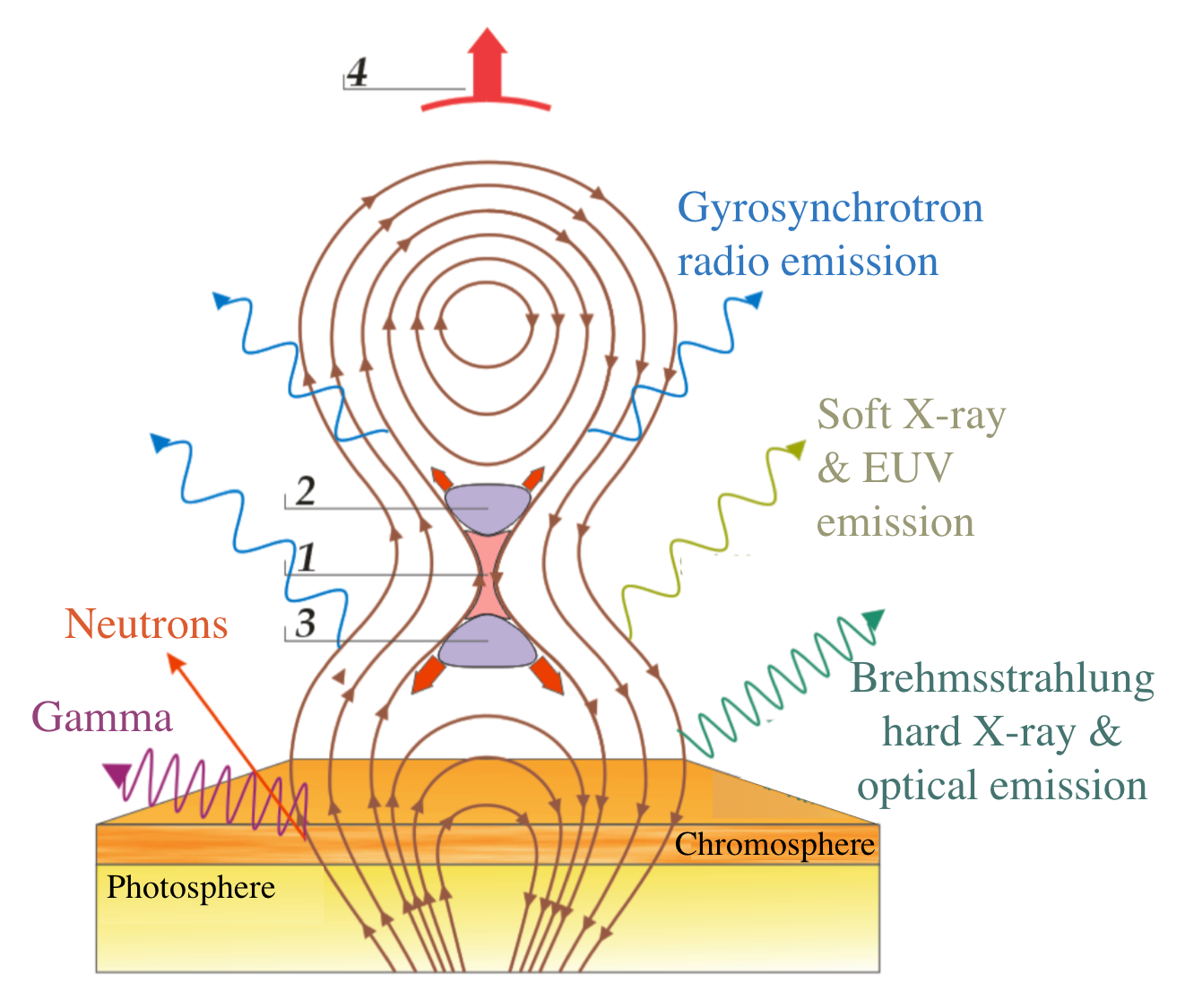} }
    \end{center}
    \caption{Schematic of solar flare with a CME showing (1) magnetic reconnection, (2) and (3) - regions of particle acceleration above and below the reconnection region, and (4) a CME shock front. Adapted from \cite{Bazilevskaya2017}. }
    \label{fig:solarflare}
\end{figure}

Solar flares are often accompanied by ejections of mass (coronal mass ejections, or CMEs), and the probability of a CME increases with increasing flare energy \citep{Yashiro2005,Wang2007}.  Figure~\ref{fig:solarflare} shows a cartoon illustrating the relationship between the trifecta in a solar eruptive event: the flare (seen as footpoints and loops), the plasmoid lifting off to become the CME, and the open field lines along which solar energetic particles (SEPs) escape to interplanetary space. CMEs are geoeffective in the solar system and the primary source of space weather, driving various authors to consider the “space weather habitable zone” (e.g., \citealt{Airapetian2017}) as the more meaningful compared to the liquid water habitable zone \citep{Kasting1993} in our quest to detect and characterize biology on other worlds. CMEs from young stars could also dominate their rotational evolution or spin-down as they carry away angular momentum from the star \citep{Cranmer2017}.
\medskip

Expectations for a one-to-one relationship between flares and CMEs, as observed in solar eruptive events, have not been borne out with recent constraints for stellar flares \citep{Osten2015,Odert2017,Crosley2018,Crosley2018a}. The main observational methods used in the stellar context are searches for blueshifts from the chromospheric H$\alpha$ emission line during a flare, long wavelength Type II radio burst signature originating from a CME, and absorption dimming in X-rays \citep{Moschou2017}. These diagnostics each have specific pros and cons: velocity signatures are expected to occur during the flare as well as any putative ejection of mass; radio observations may not be sensitive enough to detect small ejections; increases in hydrogen column density during stellar X-ray flares is not ubiquitous; and most signatures are dependent on the line-of-sight. 
\medskip

Whereas flare energy manifests in the number and frequency of emitted photons, CME energy is kinematic: cumulative mass and speed of charged particles. Fundamentally, flares and CMEs represent two mechanisms by which considerable pent-up coronal magnetic field energy can be rapidly released. For the Sun, we are able to observe CMEs directly with coronagraphs and \textit{in situ} measurements. These traditional methods are not feasible for stellar CMEs in the near future. However, a number of other techniques for detecting solar CMEs have been developed, and \cite{Harra2016} determined that coronal dimming, the dimming of emission lines sampling the quiet corona (10$^{6}$-10$^{6.4}$ K for the Sun), is the only feature consistently associated with CMEs.
\medskip

\textbf{Coronal dimming as a method for CME characterization.} The association between solar CME kinematics and coronal dimming has been established for the Sun using spatially resolved EUV images (e.g., \citealt{Aschwanden2009}) and irradiance spectroscopy (e.g., \citealt{Mason2014,Mason2016}). The transient void left behind in the corona as a CME departs results in a flux dimming \citep{Sterling1997,Harrison2003,Aschwanden2009,Reinard2009,Dissauer2018}. Coronal dimming of the Sun due to CMEs is observed in ``Sun-as-a-star" (irradiance) spectra as 1-5\% flux decreases in hot  emission lines like Fe IX (171 \AA) and Fe XII (195 \AA) on timescales of minutes to hours \citep{Woods2011,Mason2014,Mason2016}. The dimming light curve's characteristics have been used to quantify the kinetic energy of solar CMEs. The dimming depth is proportional to the CME mass and the dimming slope is proportional to the CME speed. These empirical relationships have been validated recently by CME models \citep{Jin2017} and can be adapted and applied to the physical parameters of  non-Sun-like stars. Solar studies have found that the emission lines that are sensitive to dimming are those around the ambient coronal temperature ($\sim$10$^6$ K for the Sun; see \citealt{Johnstone2015} for measurements and estimates for other stars). Recent modeling by M. Jin (2019, in preparation) indicates that this still holds true for M dwarfs: dimming-sensitive emission lines in M dwarfs should be those that form at the corresponding ambient coronal temperatures of $\sim$10$^7$ K. EUV spectra sampling a wide range of emission lines will be necessary to detect dimming from stars with a variety of coronal temperatures.

\section*{Conclusions}
An EUV-capable observatory is required to fully characterize stars with chromospheres, transition regions, and coronae. Exploring stellar EUV emission in the time domain will allow for studies of the complex magnetic field structures of stellar atmospheres and how the structures depend on stellar parameters. The EUV also offers the clearest path toward detecting and characterizing coronal mass ejections on other stars \citep{Harra2016}, which has important implications for exoplanet atmosphere stability and stellar rotation evolution.

\pagebreak

\bibliographystyle{proposal}
\bibliography{main.bbl}

\end{document}